\documentstyle[sprocl,psfig]{article}

\def\Journal#1#2#3#4{{#1} {\bf #2}, #3 (#4)}

\def\NPA{{\em Nucl. Phys.} A}
\def\PLB{{\em Phys. Lett.}  B}
\def\PRL{\em Phys. Rev. Lett.}
\def\PRC{{\em Phys. Rev.} C}

\begin{document}

\title{``DAY 1'' PHYSICS AT RHIC: PREDICTIONS  FROM  RQMD}

\author{H.~SORGE}

\address{Department of Physics \& Astronomy \\ SUNY at Stony Brook,
Stony Brook, NY 11794, USA\\E-mail: Heinz.Sorge@sunysb.edu}

\maketitle\abstracts{ 
I discuss predictions  based on the transport theoretical approach 
``relativistic quantum molecular dynamics''.  They can be tested rather soon 
by the upcoming experiments at the Relativistic Heavy Ion Collider at BNL 
(RHIC). Here I focus on the question whether hadronic observables are sensitive 
to  the Equation of  State (EOS) in the ultradense matter. A new version of
RQMD  has been developed recently in which the EOS can be varied, for instance  
like the one observed in lattice gauge studies. It is found that --  somewhat 
paradoxically --   the partial transparency of the two nuclei provides new 
avenues to assess the influence of baryon number in hot matter.  One of the 
signals which are most sensitive to the dynamics is the directed flow of   
nucleons. As a function of rapidity it changes its direction three times. 
Surprisingly, even one of the most simple observables -- the average transverse 
momenta or slopes -- displays some significant sensitivity to the EOS in the 
phase transition region between hadronic and quark matter.
}

\section{Introduction}

With experiments at the Relativistic Heavy Ion Collider at BNL (RHIC)
upcoming soon, heavy ion physics enters a new stage. It is expected
that the energy densities which may be created are favorable for the
creation of so-called quark-gluon plasma (QGP). Observation of 
this state and the transition
between QGP and hadronic matter is the foremost goal of the ultrarelativistic
heavy-ion program started more than a decade ago at CERN and at BNL with
fixed-target experiments (beam energies up to 200~AGeV).
Indications from these experiments show some clear deviations from ``linear''
$pp$ and $pA$ extrapolations  which are attributable to interactions
of the strongly interacting  matter during its dense stages 
(charmonium suppression, excess dileptons, transverse flows and strange
anti-baryon enhancement) \cite{UHe99}.
In view of these promising signals it is very important that 
Au(100AGeV) on Au(100AGeV) collisions as planned  at RHIC will provide
the opportunity to study a system which -- initially -- may be even deeper
in the quark-gluon phase than at the lower beam energies. 
Of course, with lack of data it is not easy to estimate the 
initial energy density. From an extrapolation based on the ratio of
produced charged particle 
densities at midrapidity in $pp$ and $\bar{p}p$ respectively
(2.4:1.5) one would expect a sixty percent increase. 
Such a naive scaling with $dN/dy$ of the elementary system is actually 
borne out by the calculations with the RQMD model  \cite{Sor95}. 
In Fig.~1 the  time evolution of
   energy  density  (together with  the pressure)
\begin{figure}[htp]
\vspace{-2.2cm}

\centerline{\hbox{
\psfig{figure=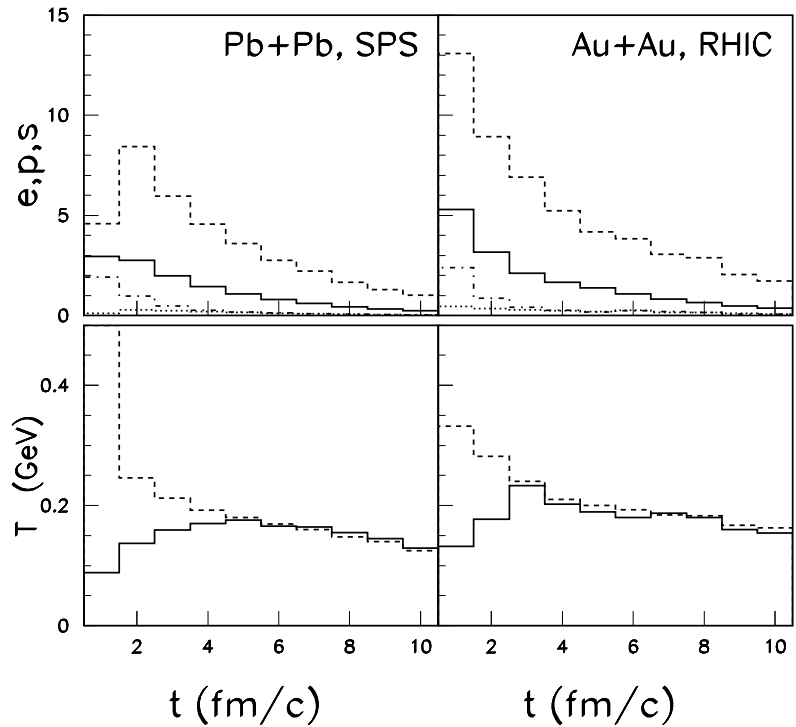,width=13cm,height=13cm}}}

\vspace{-1.0cm}
\caption
[
 ]
{
   Comparison between  Au(100AGeV) on Au(100AGeV) at RHIC
   and Pb(158AGeV) on Pb at SPS.
   The RQMD calculations were done utilizing a resonance matter EOS.
   Upper panel shows 
  time evolution of pressure $p$ (dashed-dotted line= longitudinal,
  dotted  line= transverse), 
   energy  density $e$ (solid line)  
  and entropy density $s$  (dashed  line)
  in the collision center.
  The lower panel  shows the ``temperature'' $T$ 
    extracted from the diagonal components of the energy-momentum tensor
   and the particle density. It coincides with the real temperature only
   after the pressure becomes isotropic and longitudinal $T$ values
   (dashed  line) coincides with the transverse values (solid line). 
 \label{eps}
}
\end{figure}
 is compared
  between
 heavy-ion reactions at SPS and at RHIC energy. It shows that the
 RHIC collisions produce the denser and hotter system. 
 From the lower panel of the Figure it may be read off that the thermalization
 proceeds faster at RHIC (after 3fm/c versus 5fm/c for the SPS energy).
 The dynamical reason for faster equilibration is that the role of the
 ingoing baryons is diminished with higher beam energy. 
 Since these are
 transported from the original rapidities to midrapidity,
  they  tend to cause stronger non-equilibrium effects.
 In contrast,
 produced particles are locally correlated tighter  in rapidity. Fig.~1 also
 shows that after equilibration
 the system stays at temperatures above $T_c \approx $ 160 MeV
 in the center for 6~fm/c.

On the theoretical side, the ``Achilles heel'' of the search for the
QGP and the transition is an uncertainty what the properties
of the QGP are. Smilga among others has pointed out that the physics of the state
above and close to $T_c$ is a theoretical ``no man's land'' \cite{Smi99}. Neither
perturbative QCD nor interactions between (quasi-)hadronic
states appear justified for a description. If one can write down no equations
and give no numbers
the question whether one has actually ``seen''  a QGP may deteriorate into an
exercise in semantics. One way out is to resort to models of QGP and the
phase transition. For instance,  $J/\Psi$ suppression
was initially predicted by Matsui and Satz as due to static color screening
between heavy quarks in the plasma \cite{MaS86}. 
Lateron,  $J/\Psi$ suppression  was indeed  observed 
in Pb(158AGeV) on Pb collisions (and before in lighter systems).
The model predicts that the $\chi $ states which are orbital
excitations feeding into $J/\Psi$ are even more suppressed than the
$\Psi '$  -- in conflict with the observations. Do we infer that the model
fails for heavy-ion
 reactions or that no QGP has been produced at CERN-SPS energies?
 
\section{Radial and directed collective flow at RHIC}

 Another way out of the difficulty to find signatures of the QGP 
 is to design ``robust'' observables which do not rely
 on detailed assumptions. Here one class of observables
 -- real photons or dileptons emerging from virtual photons -- can be
 primarily viewed as a tool to extract the highest temperature
 from the photon and dilepton spectra. Unlike hadrons, even early
 produced photons 
 leave the system without much disturbance.
 The other main tool to store information from the early stages is
 collective transverse flow. Essentially, the flow velocities can be
 represented as a time integral over the forces acting on the matter
 during evolution. Close to equlibrium, these forces are determined by
 the pressure gradients (and, unfortunately, some non-ideal properties
 of the matter like viscosities).  The pressure is a fundamental
 thermodynamic quantity and shows a rather characteristic behaviour in the
 (phase?) transition region between hadronic and quark matter. The EOS
 displays a ``softest point'' leading to a minimum of the expansion velocity
  \cite{MHS95}.
  In order to infer information on the acceleration
 history one needs to de-convolute the accumulated flow. Various
 techniques are available such as different types of flow which
 have different sensitivity to the early reaction stages (elliptical
  and directed versus radial flow) \cite{Sor97a,Sor97b} 
 and utilizing different
 particle species. In particular,  multi-strange hadrons have typically
 smaller cross sections and thus decouple from the system at earlier 
 times   \cite{HSX98,Dum99}.

 An obvious question for collisions at RHIC energy is whether the
 physics of the ``softest point'' is not better studied at lower
  energies.  Presumably, the EOS  is softest at  energy densities
  around 1~GeV/$fm^3$. 
  Indeed, various questions are currently being pursued   like
   its influence on elliptical flow or the chemistry in 
  $AA$ collisions with beam energies as low as
   2~AGeV. On the other side, it is one thing to infer ``softening''
   of the EOS from data   \cite{E895}.
   It is quite more convincing evidence  for the ``phase transition''
   (which may in fact be just a smooth cross-over)
   to observe the re-hardening
   of the EOS at larger energy densities as well. 
  From lattice results and general considerations an asymptotic 
  energy density --
   pressure relation like $p\sim e/3$ is inferred, characteristic for
  an ideal gas of massless quarks and gluons.
  How can we observe whether the system's evolution is partially characterized
    by such a hard EOS?
  The flow develops as an integral over time whose
  characteristic scale is set by the transverse size of the system.
   The hardness of the EOS  at times around $r_{tr}/c$ is more relevant
   concerning flow development than the degree of equilibration and the
   resulting maximum energy densities just after impact ($t\le 1 fm/c$).
   A look at Fig.~1 will tell that RHIC may be in a better position than
   the SPS program for studying the EOS above $T_c$. 
  Typically,  results of dynamical  calculations (transport  or
  hydrodynamical) agree that there is not much sensitivity to the
   re-hardening of the EOS up to the highest SPS energy 
    \cite{JSo96,Sch98}. 
  Only recently
  it was suggested that perhaps the  ``kinky'' 
  centrality dependence of elliptical flow
  may provide evidence for a valley-type structure of $p/e$,
  i.e. the softening and re-hardening of the EOS   \cite{Sor99}. 
  Of course, the  smaller characteristic size in semi-peripheral collisions
  is helpful. It is much easier to
  observe  early-time phenomena in small  systems.
  Several recent studies find that elliptical flow is a useful
  and measurable observable at RHIC as well  \cite{Zha99,TeS99,SPV99}. 
   
\begin{figure}[htp]
\vspace{-2.2cm}

\centerline{\hbox{
\psfig{figure=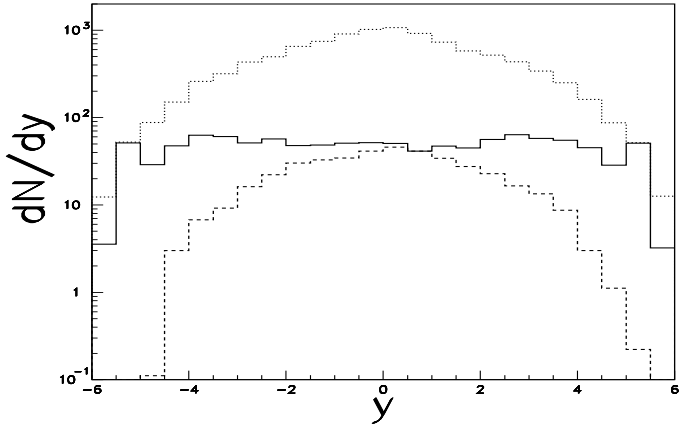,width=10cm,height=8cm}}}

\vspace{-1.0cm}
\caption
[
 ]
{
  RQMD result for the rapidity distribution of nucleons, anti-nucleons
  and pions  in Au(100AGeV) on Au(100AGeV) collisions at $b=3$~fm.
 \label{barmesdndy}
}
\end{figure}

 Fig.~2 displays the  rapidity distributions of nucleons, anti-nucleons
 and pions
 in Au(100AGeV) on Au(100AGeV) collisions at $b=3$~fm as calculated from RQMD. 
 According to the calculations the in-going baryons
 end up mostly in-between midrapidity and the original rapidities. 
 In RQMD and other sensible models of baryon stopping the colliding
 nuclei become more transparent to each other with higher beam energies.
 It is usually assumed 
  that the fast valence quarks are  mere spectators
  in the soft reactions  which are initiated by the soft glue 
  $x_F \rightarrow 0$.  Thus
 leading particles obey Feynman scaling which translates into a scaling
  rapidity loss distribution at high energies. For baryons 
   the average loss is $\Delta y \sim -2$ in central heavy-ion collisions. 
 It is one of the intriguing aspects of the  baryon stopping that
 shift of baryon number and valence quarks may not coincide. 
 (It has not been checked for real collisions, but in the model they do not.
  ``Net-valence'' distributions constructed from e.g.
   net kaon distributions show more transparency than the
  net-baryon distributions.)
  In a model
 like RQMD the ``junction'' which connects the $N_c=3$ valence quarks
 via ``strings'' is an independent dynamical object \cite{Sor95}.
 If all valence quarks are stripped off the junction, the emerging baryon
 is stopped more than the valence quarks. 
 (There have been other talks in this meeting devoted to the subject
  of baryon stopping, e.g. Huang's talk.)

\begin{figure}[htp]
\vspace{-1.2cm}

\centerline{\hbox{
\psfig{figure=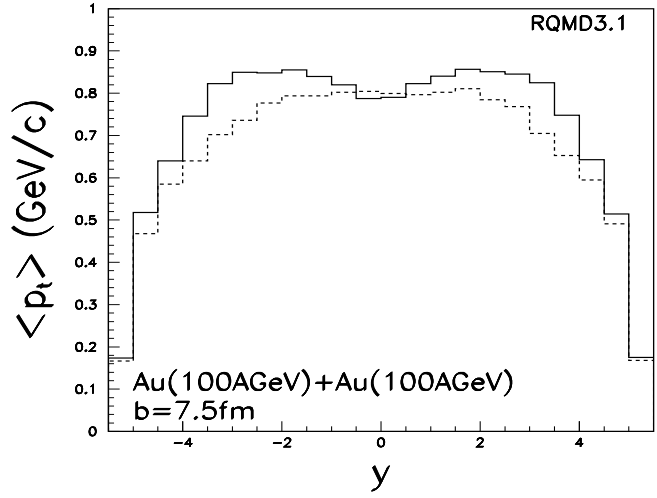,width=10cm,height=9cm}}}

\vspace{-1.0cm}
\caption
[
 ]
{
  Average transverse momentum  of nucleons
   as a function of $y$: 
   EOS w. 1st order phase transition (solid line), resonance gas
   (dashed line).
 \label{meanpt}
}
\end{figure}

\begin{figure}[htp]
\vspace{-0.6cm}

\centerline{\hbox{
\psfig{figure=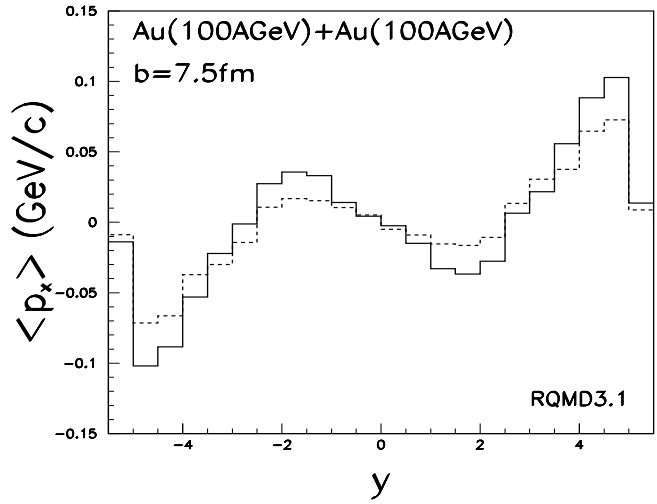,width=10cm,height=9cm}}}

\vspace{-1.0cm}
\caption
[
 ]
{
  Average transverse momentum component of nucleons
  in the reaction plane as a function of $y$ (directed flow): 
   EOS w. 1st order phase transition (solid line), resonance gas
   (dashed line).
 \label{dirflow}
}
\end{figure}
   
A small value of net-baryon rapidity density leads to small spatial
net-baryon densities in the central region. However,
with the baryon densities peaked off midrapidity
a strong gradient of baryon density with increasing rapidity is
created.
Using the idealizing equilibrium assumption, going away from $y$=0
 corresponds to walking to finite $\mu _B$ in the $T$-$\mu _B$ plane.
It allows to obtain information on the phase diagram of QCD by
correlating baryon number in different 
rapidity windows to observables, possibly even event-by-event. 
Fig.~3 shows that the role of baryons may have striking consequences
for ``radial'' transverse flow (the dominating isotropic component).
In the figure  the average transverse momentum of
protons is shown as a function of rapidity calculated with RQMD.
The average $p_t$ of nucleons serves as an experimentally accessible
proxy for the not directly observable radial expansion. 
Two results are compared, one obtained with a resonance matter EOS
($p/e\sim 6$), the other based on an EOS with 1st order phase transition
 \cite {Sor99}.
It was calculated in
a  bag-type model of quarks and gluons with temperature-dependent
masses and ``bag constant''  which agrees well with lattice
data (small latent heat).
We observe that the flow (average transverse momentum)
is  larger at non-zero rapidity
-- contrary to any naive ideas about ``more action in the center''
and observations at fixed-target energies. 
At midrapidity  both EOSs lead to the same hardness of the 
transverse momenta. Additional hardness of the EOS in the quark-gluon phase
is canceled by the  softness in the ``mixed'' phase
compared to the structureless resonance matter EOS.
The very good agreement of  resonance gas
dynamics with data as observed at SPS energy (158--200AGeV)  \cite{Sor97b}
may have been accidental to some extent, by hitting just the
right average pressure. 
At RHIC -- away from midrapidity -- adding  baryons 
destroys the balance between additional  hardness and softness 
increasing the total amount of flow. 

Another handle on the EOS in the ultra-dense matter is provided by
the so-called directed flow (a ``kick'' of matter in the reaction
plane which is balanced by the corresponding ``anti-kick'' at opposite
rapidity. Fig.~4 displays the calculated
  average transverse momentum component of nucleons
  in the reaction plane as a function of $y$. 
  From fixed-target energies the S shape of $p_x(y)$ is well-known. 
 For collider energies RQMD predicts 
  a more complicated structure \cite{SSV99}. 
  The two wings of the ``S'' are still present
  close to the projectile and target rapidity respectively.
  However,  the ``S'' is broken into two pieces.
  Nucleons at more  central rapidities tend to move into
  the opposite direction from their fellow nucleons in the same hemisphere.
  The dynamical reason for this unorthodox behaviour is the
  inhomogeneous transverse source distribution of the ingoing baryons
   after the initial impact. 
  The large rapidity gap between projectile or  target nucleus
  and central rapidity region can only be overcome if nucleons pass
  through as much nuclear matter as possible. That is just the side
  opposite to the spectators comoving in their direction
  which define the standard S shaped configuration. 
  Furthermore, the large
   gap of five units in rapidity means that typically nucleons
   stay in their rapidity hemisphere even if they are close to midrapidity.
   Therefore  ingoing  baryons  
   find themselves on  opposite sides of the reaction zone
   depending on the sign of their rapidity (in the center of mass).
   The subsequently developing transverse flow converts the initial
   directional  asymmetry in the reaction plane into  a directed flow signal. 
 In Fig.~4 
  the results for the        
   EOS with 1st order phase transition are compared to the
  results with the resonance gas  EOS. 
  It becomes apparent from a comparison with the results for average
   transverse momenta that the transverse momentum component
 in the reaction plane
  is clearly more sensitive to the EOS.
  Uniformly, the directed flow is more pronounced with the QGP based EOS
  which points to a larger sensitivity to the early evolution for this
   observable. 
 
\section{Conclusions and Disclaimer}

Of course, some caveats are in order about the calculations. 
The QGP based EOS has been implemented in the RQMD model as $p(e)$, i.e. 
baryons  contribute only implicitly -- via their energy. 
This prescription does not need to be true. 
We do not know too much about the EOS of QCD at finite
temperature and baryon density. Basically, there are two schools of
thought, one empirically oriented and one  studying simple
models (NJL etc.) which  resemble QCD. 
 One can constrain the baryon density dependence of the EOS
from heavy-ion reactions 
at  beam energies down to very low values ($\sim $1--10~AGeV,
 the AGS region). 
This has been done in the past by many people. The debate
is not closed yet and oscillates between  preference for a ``hard''
versus ``soft'' EOS (with a remarkable come-back for the former)
\cite{SSO99,BAL99}.
A recent variation of the  ``hard versus soft'' debate
is that perhaps the EOS is  hard first but  softens with increasing baryon
density \cite{PDA98}. 
It should be noted that the scale defining ``softness''
is very different from the baryon-number suppressed
high-energy (SPS) scale. On the low-energy (=baryon-rich) scale
a resonance gas EOS is actually ultrasoft and ruled out by plenty of data
in the energy range 1--15~AGeV  \cite{Efr98}.
The bottom line from HI physics is therefore that baryons  make the EOS
more repulsive compared to  baryon-free matter at same energy density.
An interesting consequence could be that a
1st order(-like)  transition at $\mu_B$=0  is ``killed''
by adding baryon number.
On the other side, based on NJL-type models we expect
just the opposite, an almost 2nd order transition at
 $\mu_B$=0 (almost because of the nonzero masses of light flavor quarks)
and a 1st order transition at $T$=0 and finite $\mu_B$. 
An interesting consequence would be that the 1st order transition
line ends in a tri-critical point somewhere in the
$T-\mu_B$ plane \cite{SRS98}. Of course, 
having in mind that nuclear matter exerts more pressure than pions 
it may be just the other way around. 
In that case  a tricritical point would be connected to $T_c$
at  $\mu_B$=0 via a  1st order transition line. 
More detailed studies of this possibility are currently under way 
\cite{LOI99}.
Lattice calculations are
not conclusive yet about the nature of the
transition at finite $T$. The mass of the strange quark  is the decisive
factor.  
Yet another phase transition  at finite baryon density may be associated
with a super-conducting phase of QCD in which color symmetry is spontaneously
broken \cite{RSS98,ARW98}. However,  this phase is probably not accessible
in heavy ion collisions, because they are too hot. It has  
been claimed that the observed persistence of magnetic fields in neutron
stars rules out such phases and the large associated gaps 
even at $T$=0 for nuclear densities
in the range of up to 8~$\rho _0$ \cite{Hsu99}. 
In any case,   some clarifications of the QCD phase diagram
 at high $T$ and -- depending on rapidity window -- zero or nonzero $\mu _B$ 
 may be within reach with the upcoming RHIC experiments.

\section*{Acknowledgments}
This work has been
supported by DOE grant No. DE-FG02-88ER40388.

\end{document}